\documentclass[twocolumn,aps,prd,preprintnumbers,nofootinbib]{revtex4}
\usepackage[dvipdfmx]{graphicx}
\usepackage{bm,latexsym,amsmath,amssymb,amsfonts,color}

\usepackage[normalem]{ulem}

\begin{document} 

\title{\uline{}New definition of wormhole throat} 

\author{Yoshimune Tomikawa${}^1$, Keisuke Izumi${}^2$, Tetsuya Shiromizu${}^{1,3}$}

\affiliation{${}^1$Department of Mathematics, Nagoya University, Nagoya 464-8602, Japan}
\affiliation{${}^2$Leung Center for Cosmology and Particle Astrophysics, National Taiwan University, Taipei 10617, Taiwan}
\affiliation{${}^3$Kobayashi-Maskawa Institute, Nagoya University, Nagoya 464-8602, Japan}
\begin{abstract}
We present a new definition of the wormhole throat including the flare-out condition and 
the traversability for  general dynamical spacetimes in terms of null geodesic congruences. 
We will examine our definition for some examples and see advantages compared to the others. 
\end{abstract}
\maketitle

\section{Introduction}

Wormhole is one of the interesting objects in general relativity 
\cite{M-T,M-T-Y,visser,lobo2007}. Although we have a vague image of wormhole, there is 
no universal definition which can work for general situations. To discuss 
wormhole we have to specify the throat, the flare-out condition and so on. There are 
some proposals \cite{M-T,H-V1998prl,hayward1999,M-H-C}. As far as we know, for static 
and spherical symmetric cases, the definition of the throat was firstly given in Ref. \cite{M-T}. 
Therein, the static slices are embedded to the Euclid space to see the throat structure. 
Since the symmetries are used for the definition of wormhole,
it is clear that this proposal is not applicable for dynamical or non-spherical symmetric cases. 

The issue about the extension of the concept of wormhole to general spacetimes has 
been already addressed in Refs. \cite{H-V1998prl, hayward1999} 
(see also Refs. \cite{H-V1997, H-V1998prd, H-V1999, hayward2009, M-H-C}). 
In Refs. \cite{H-V1998prl, hayward1999}, 
using null geodesic congruences, 
the wormhole throat is defined as the minimal surface on the null hypersurfaces, i.e.
the trapping horizon~\cite{hayward1994}. 
We know that, in their definition, some exotic matters are required for static wormholes 
\cite{M-T, M-T-Y, H-V1997, hayward2009, V-K-D} and for dynamical ones \cite{H-V1998prl, H-V1998prd, H-V1999} 
without any singularities. Nevertheless, our Universe has the initial singularity, and 
cosmological wormhole solutions with the initial singularity, where two 
Friedmann-Lema\^{\i}tre-Robertson-Walker (FLRW) universes are connected, were constructed 
without any exotic matters~\cite{M-H-C}. 
Since these solutions do not meet the definition of Refs.~\cite{H-V1998prl, hayward1999},
the authors proposed an alternative definition focusing on spherical symmetric cases; 
the wormhole throat is the minimal surface on spacelike hypersurfaces~\cite{M-H-C}. 
However, this definition strongly depends on which spacelike hypersurfaces we take and, 
because of this dependence, even de Sitter  
and FLRW spacetimes are categorized into wormhole. 

In this paper, we would propose a new definition of the wormhole throat which is better 
suited for the intuitive image of wormhole. Our definition seems to be a hybrid one 
of null-hypersurface-based definition \cite{H-V1998prl, hayward1999} and spacelike hypersurface 
one \cite{M-H-C}. We describe the throat in terms of the expansion rate of null geodesic 
congruences on a kind of spacelike hypersurface. 

The other parts of this paper are organized as follows.
In Sec. \ref{setup}, we introduce a new definition of the wormhole throat and discuss 
some general features. In Sec. \ref{example}, we consider several examples to see if our 
definition can work well. Finally, we give a summary in Sec. \ref{summary}.

\section{New definition} \label{setup}

In this section, we propose a new definition of the wormhole throat with the 
flare-out condition and the traversability. 
We also discuss some general features.

We consider a codimension two spacelike compact surface $S$ and future directed outgoing/ingoing null geodesic 
congruences with the affine parameter $\lambda_\pm$ emanating from $S$. Then we define the null 
expansion rate $\theta_\pm$ and we introduce the following quantities 
\begin{equation}
k:=\theta_+-\theta_- \label{kpm}
\end{equation}
and
\begin{equation}
\bar k:=\theta_++\theta_-. 
\end{equation}
Defining the following two vectors 
\begin{equation}
r^a:=(\partial_+-\partial_-)^a
\end{equation}
and
\begin{equation}
t^a:=(\partial_++\partial_-)^a,
\end{equation}
where $\partial_\pm:=\partial_{\lambda_\pm}$, $k$ and $\bar k$ are rewritten as 
\begin{equation}
k=r^a \nabla_a \ln {\sqrt {h}},~~\bar k=t^a \nabla_a \ln  {\sqrt {h}}. 
\end{equation}
In the above $h$ is the determinant of the induced metric of the codimension two surface $S$. 
Although we cannot assume that the affine parameter $\lambda_\pm$ emanating from $S$ provides 
us the global coordinate for spacetimes in general, we can have a quasi-local null 
coordinate system $\tilde \lambda_\pm$ such that it coincides with $\lambda_\pm$ when it 
crosses $S$. 

Now we define the throat as the codimension two surface such that 
\begin{eqnarray}
k|_S=0 \label{mins}
\end{eqnarray}
holds and  the following flare-out condition
\begin{eqnarray}
r^a \nabla_a k|_S>0 \label{foc}
\end{eqnarray}
is satisfied. We emphasize that, by fixing the coordinate locally with $\tilde \lambda_{\pm}$, 
there is no ambiguity of the spatial derivative $r^a \nabla_a$.
To introduce the concept of the traversability for the wormhole we consider 
the time sequence of the throat. We say that the wormhole is traversable if the 
tangent vector of the sequence of the throat, which is normal to $S$, 
is timelike. Moreover, if there are some event horizons in the region that satisfies 
Eq. (\ref{mins}) and inequality (\ref{foc}), we exclude the inside of 
the event horizon from the definition of the throat. This is because 
travelers cannot come back to the same region from black hole. 

Let us look at general properties of our definition of the wormhole.
From the condition of Eq. (\ref{mins}) we have 
\begin{eqnarray}
\theta_+|_S=\theta_-|_S.
\end{eqnarray}
When $\theta_+|_S=\theta_-|_S<0 ~(>0)$, it means the existence 
of the future(past) trapped surface. Then, if the null energy condition holds, 
the singularity theorem implies the presence of singularity 
in the future(past) \cite{wald}.  
With the energy condition, assuming the cosmic censorship conjecture to be held, 
the future trapped region is always inside the event horizon \cite{wald}, and  
thus it is excluded in our definition of wormhole. 
Meanwhile, with $\theta_+|_S=\theta_-|_S>0$,
the singularity theorem predicts the past singularity but we do not exclude the past trapped region.
The realisation of $\theta_-|_S>0$ 
will be easy in expanding universe and the past singularity may be unified 
to the initial one. That is, there is a room to construct a dynamical 
wormhole in the cosmological context keeping the energy condition. 

Let $z^a$ to be the tangent vector of the time sequence of the throat which is 
normal to $S$. 
Since it is timelike, we can write it as 
$z^a=\alpha (\partial_+)^a+\beta (\partial_-)^a $ with $\alpha,\beta>0$. 
Along the time sequence, 
\begin{eqnarray}
z^a \nabla_a k|_S=0 \label{tvc}
\end{eqnarray}
holds and this gives 
\begin{eqnarray}
\partial_- k|_S=-\frac{\alpha}{\beta}\partial_+k|_S. 
\end{eqnarray}
Then, $r^a \nabla_a k|_S$ becomes 
\begin{eqnarray}
r^a \nabla_a k|_S& = & 
\Bigl(1+\frac{\alpha}{\beta} \Bigr) \partial_+k|_S \nonumber \\
& = & \Bigl(1+\frac{\alpha}{\beta} \Bigr)(\partial_+ \theta_+-\partial_+ \theta_- )|_S>0.
\end{eqnarray}
If the null energy condition is satisfied in $D$-dimensional spacetimes, the Raychaudhuri 
equation tells us 
\begin{eqnarray}
\partial_+ \theta_+ =-\frac{1}{D-2}\theta_+^2 -\sigma_{+ab}\sigma_+^{ab}-R_{ab}n^an^b \leq 0,
\end{eqnarray}
where $\sigma_{+ab}$ is the shear and $n^a$ is the tangent vector of null geodesics. 
Here we used the fact that the null geodesic congruences are normal to the throat, that is, 
the rotation of the congruence vanishes. 
Therefore, 
\begin{eqnarray}
\partial_+\theta_-|_S<0
\end{eqnarray}
is required at least for the presence of the throat. 

It is nice to have other general features. 
Since we have an equality
\begin{eqnarray}
r^a \nabla_a k+t^a \nabla_a \bar k=2(\partial_+ \theta_++\partial_- \theta_-), 
\end{eqnarray}
the Raychaudhuri equation with the null energy condition show us 
\begin{eqnarray}
r^a \nabla_a k+t^a \nabla_a \bar k \leq 0. 
\end{eqnarray}
In particular, the flare-out condition is not satisfied, $r^a \nabla_a k \leq 0$, for a sort of static case 
of $t^a \nabla_a \bar k=0$ 
as long as the null energy condition is satisfied. This is a simple confirmation of 
well-known fact.

\section{examples} \label{example}

Let us examine our definition in four dimensional spacetimes with symmetries 
including the spherical symmetry. 

\subsection{General scheme} \label{conformal}

In the null coordinate, the metric of a spherically symmetric spacetime is generically written as 
\begin{eqnarray}
ds^2 = -a^2(u,v) dudv +R^2(u,v) d\Omega^2_2, \label{met1} 
\end{eqnarray}
where $d\Omega^2_2$ is the metric of the unit 2-sphere. 
The throat is supposed to be located at a two surface specified by $u=u_0, v=v_0$.

The radial null geodesic will be on $u$ or $v=$constant lines. Let us consider 
the geodesic on $v=v_0$ which follows the geodesic equation 
\begin{equation}
\frac{d^2u}{d\lambda_u^2}+2\frac{\partial_u a}{a}\Bigl(\frac{du}{d\lambda_u} \Bigr)^2=0.
\end{equation}
In a formal way, we can solve the above as 
\begin{equation}
\lambda_u=C_u^{-1}\int^u a^2 (u')du'=:U,
\end{equation}
where $a(u):=a(u,v_0)$. $C_u$ is the positive integration constant and we choose $\lambda_u$ such that 
$du/d\lambda_u >0$. In the same way, for the 
geodesic on $u=u_0$, we have 
\begin{equation}
\lambda_v=C_v^{-1}\int^v a^2 (v')dv'=:V,
\end{equation}
where $a(v):=a(u_0,v)$. $C_v$ is the positive integration constant and we choose $\lambda_v$ such that 
$dv/d\lambda_v >0$. 
Employing $U,V$ as new coordinates,  the metric (\ref{met1}) is rewritten as
\begin{eqnarray}
ds^2 =-C_u C_v \frac{a^2(u,v)}{a^2(u) a^2(v)} dUdV +R^2(u,v) d\Omega^2_2. \label{nullco}
\end{eqnarray}
We would stress that $U$ ($V$) is the affine parameter on $v_0~(u_0)=$constant geodesic. 

The null expansion rate $\theta_U,\theta_V$ are calculated to be 
\begin{eqnarray}
\theta_{U} =\theta_- =\dfrac{2}{R} \partial_U R,~~ \theta_{V} =\theta_+=\dfrac{2}{R} \partial_{V} R.
\end{eqnarray}
So $k$ defined by Eq. (\ref{kpm}) becomes 
\begin{eqnarray}
k &=& \dfrac{2}{R} (\partial_V -\partial_U )R \nonumber \\
&=& \dfrac{2}{R} (C_v a^{-2}(v) \partial_v-C_u a^{-2}(u) \partial_u)R. 
\end{eqnarray}
On the throat $S$, $k$ is supposed to vanish and then we have 
\begin{eqnarray}
C_u \partial_u R|_S =C_v \partial_v R|_S. \label{conk}
\end{eqnarray}
In the above, we used the fact of $a(u_0)=a(v_0)=:a_0$.

We also need to check the flare-out condition (\ref{foc}) and the traversability (\ref{tvc}), 
which are written with the coordinate (\ref{nullco}) as
\begin{eqnarray}
&& r^a \nabla_a k|_S \nonumber \\
&& =(\partial_V -\partial_U) k|_S \nonumber \\
&& = \dfrac{2}{a_0^4 R} \left[ -2 C_v^2 \partial_v \ln a(v) \partial_v R 
-2 C_u^2 \partial_u \ln a(u) \partial_u R \right. \nonumber \\
& & ~~~~~~~~~~\left. + C_v^2 \partial_v^2 R -2 C_u C_v \partial_u 
\partial_v R +C_u^2 \partial_u^2 R \right] \big|_S \nonumber \\ 
&& >0\label{conr}
\end{eqnarray}
and
\begin{eqnarray}
&& z^a \nabla_a k|_S \nonumber \\
&& =(\alpha \partial_V +\beta \partial_U) k|_S \nonumber \\
&& = \dfrac{2}{a^4 R} \left[ -2\alpha C_v^2 \partial_v \ln a(v) \partial_v R 
+2\beta C_u^2 \partial_u \ln a(u) \partial_u R \right. \nonumber \\
& & ~~~~~~\left. +\alpha C_v^2 \partial_v^2 R -(\alpha -\beta) C_u C_v \partial_u 
\partial_v R -\beta C_u^2 \partial_u^2 R \right] \big|_S \nonumber \\ 
&&=0 .\label{conz} 
\end{eqnarray} 
Equalities (\ref{conk}), (\ref{conz}) and inequality (\ref{conr}) 
with the positivities of $C_u$, $C_v$, $\alpha$ and $\beta$ are 
the conditions for the wormhole in  spherically symmetric spacetimes.

\subsection{Examples} \label{ex}

In this subsection, we  look at concrete examples which include non-wormhole 
spacetimes. 

\subsubsection{Schwarzschild spacetime}

It is well-known that the throat of the Schwarzschild spacetime is not the throat of the wormhole due to the presence of the event horizon.
Nevertheless, it is nice to see the feature in terms our definition of the throat. 
To see this we  adopt the Kruskal coordinate 
\begin{eqnarray}
ds^2=\frac{4r_g^3e^{-r/r_g}}{r}(-dT^2+dX^2)+r^2 d \Omega^2_2, \label{kru}
\end{eqnarray}
where $r_g=2M$ and $M$ is the Arnowitt-Deser-Misner(ADM) mass. The coordinate transformation from 
the Kruskal to ordinal one is given by  
\begin{eqnarray}
(r/r_g-1)e^{r/r_g}=X^2-T^2
\end{eqnarray}
and
\begin{eqnarray}
T/X=\tanh (t/2r_g)
\end{eqnarray}
for $r>r_g$, or 
\begin{eqnarray}
X/T=\tanh (t/2r_g)
\end{eqnarray}
for $0<r<r_g$. 

In this case, choosing $u,v$ as $u=T-X ,v=T+X$,  $k|_S=0$ (Eq. (\ref{conk})) 
gives us 
\begin{eqnarray}
C_u (T+X)|_S =C_v (T-X)|_S. \label{kru-th}
\end{eqnarray}
This implies that the  candidate of a throat is in the region $0<r \leq r_g$ because of $C_u, C_v>0$.
In addition, inequality (\ref{conr}) and Eq. (\ref{conz}) become 
\begin{eqnarray}
r^a \nabla_a k |_S =\dfrac{4r_g^4}{a_0^4r^4} e^{-r/r_g} C_u C_v |_S >0, \label{kru-fl}
\end{eqnarray}
\begin{eqnarray}
z^a \nabla_a k |_S=\dfrac{2r_g^4}{a_0^4r^4} e^{-r/r_g} (\alpha -\beta) C_u C_v |_S =0, \label{kru-tr}
\end{eqnarray}
where $a_0^2=4(r_g^3/r) e^{-r/r_g}$. 
The flare-out condition is satisfied as expected and the 
tangent vector of the throat orbit of $\alpha =\beta$ is timelike.
However, there is the event horizon at $r=r_g$ that is the boundary of the  region satisfying Eq. (\ref{conk}) and inequality (\ref{conr}). 
Therefore,  as we have commented in Sec.~\ref{setup}, the Schwarzschild spacetime does not have the wormhole throat.

\subsubsection{de Sitter spacetime}

Next we will examine the de Sitter spacetime.  
If one elaborates the selection of 
a spacelike hypersurface and follows Maeda et al.'s definition \cite{M-H-C} 
for the wormhole throat, there is a case where the wormhole is. This is because 
Maeda et al.'s definition is not slightly appropriate. Meanwhile our definition 
excludes this case. 

In the flat chart, the metric of the de Sitter spacetime is given by 
\begin{eqnarray}
ds^2 & = & a^2(\eta)(-d\eta^2+dr^2+r^2d\Omega_2^2) \nonumber \\
     & = & a^2(\eta)(-dudv+r^2d\Omega_2^2), \label{metde}
\end{eqnarray}
where $a(\eta)=-1/(H\eta)$, $H$ is the Hubble constant and $u=\eta -r, v=\eta +r$. 
Then Eq. (\ref{conk}) implies 
\begin{eqnarray}
C_u (Har-1)|_S =C_v (Har+1)|_S. \label{ds-th}
\end{eqnarray}
This has a solution 
\begin{eqnarray}
Har=\dfrac{C_u +C_v}{C_u -C_v} >1 ,
\end{eqnarray}
if one chooses $C_u,C_v$ satisfying $C_u>C_v$. 
This means that the throat candidate is in outside of the cosmological horizon. 

Let us see the flare-out condition (\ref{conr}).  With the metric (\ref{metde}), we have
\begin{eqnarray}
r^a \nabla_a k |_S=-\dfrac{2H^2}{a^2} C_u C_v |_S<0. \label{ds-fl}
\end{eqnarray}
This disagrees with the flare-out condition (\ref{conr}). 
Therefore,  there is no throat in the de Sitter spacetime as expected.

\subsubsection{Friedmann-Lema\^{\i}tre-Robertson-Walker(FLRW) spacetime}

Now we consider the FLRW spacetime. The metric is given by 
\begin{eqnarray}
ds^2 & = & -dt^2 +a^2(t)[ (1-kr^2)^{-1} dr^2 +r^2 d\Omega^2_2] \nonumber \\
& = & a^2(\eta)[-d\eta^2 +d\zeta^2 +r^2 d\Omega^2_2] \nonumber \\
& = & a^2(\eta)[-dudv +r^2 d\Omega^2_2],
\end{eqnarray}
where $k=-1,0,1$ depending on the spatial topology, $\eta$ is the conformal time defined by 
$d\eta=a^{-1}(t)dt$, $d\zeta=dr/{\sqrt {1-kr^2}}$ 
and $u=\eta -\zeta, v=\eta+ \zeta$.

For the FLRW spacetime, Eq. (\ref{conk}) becomes 
\begin{eqnarray}
\left.C_u (\dot{a} r-\sqrt{1-kr^2}) \right|_S =\left.C_v (\dot{a} r+\sqrt{1-kr^2}) \right|_S,
\end{eqnarray}
where $\dot{a} =da(t)/dt$. If one chooses $C_u, C_v$ satisfying $C_u>C_v$,  
the above has the solution as 
\begin{eqnarray}
H(t) \frac{a(t)r}{\sqrt{1-kr^2}}=\frac{C_u+C_v}{C_u-C_v}>1, \label{FLRW>0}
\end{eqnarray}
where $H(t):=\dot a(t)/a(t)$.
Roughly speaking, as in the de Sitter case, this means that the throat candidate is 
in outside of the cosmological horizon. 

For the current case, the flare-out condition (\ref{conr}) becomes
\begin{eqnarray}
r^a \nabla_a k|_S &=& \dfrac{2C_u C_v\left[ a\ddot{a} (1-kr^2) -\dot{a}^2 r^2 (\dot{a}^2 +k) \right]}{a^4 \left[ \dot{a}^2 r^2 -(1-kr^2) \right]} \bigg| _S \nonumber \\ &>&0.
\end{eqnarray}
This requires 
\begin{eqnarray}
\dot{a}^2 r^2 (\dot{a}^2 +k) < a \ddot{a} (1-kr^2). \label{FLRWr}
\end{eqnarray}
Together with Eq. (\ref{FLRW>0}), the above implies 
\begin{eqnarray}
\dot{a}^2 r^2 (\dot{a}^2 +k)<a\ddot{a} (1-kr^2) <a\ddot{a} \dot{a}^2 r^2. \label{F}
\end{eqnarray}
Using the Friedmann equation, it is easy to see that the inequality  
$\dot{a}^2 +k<a \ddot{a} $ obtained from inequality (\ref{F}) is 
equivalent with the violation of the null energy condition, 
\begin{eqnarray}
\rho +p<0,
\end{eqnarray}
where $\rho$ and $p$ are the energy density and the pressure of the perfect fluid, respectively. 
This is consistent with common sense.

\subsubsection{Morris-Thorne wormhole}

The Morris-Thorne wormhole, which is static and spherically symmetric, 
is often investigated \cite{M-T, M-T-Y, visser, lobo2007, hayward2009, V-K-D}.
The metric is given by
\begin{eqnarray}
ds^2 &=& -e^{2\Phi (r)} dt^2 +\left( 1-\dfrac{b(r)}{r} \right) ^{-1} dr^2 +r^2d\Omega_2^2 \nonumber \\
&=& e^{2\Phi} (-dt^2+d\zeta^2)+ r^2 d\Omega_2^2 \nonumber \\
&=& -e^{2\Phi}dudv +r^2 d\Omega_2^2,
\end{eqnarray}
where $\Phi (r), b(r)$ are functions of $r$, $\zeta$ is defined by 
$d\zeta =e^{-\Phi}dr/\sqrt{1-\frac{b}{r}}$ and $u=t-\zeta, v=t+\zeta$. 
Here we suppose that $g_{tt}=-e^{2\Phi}$ is negative and regular. 
Note that this metric is not obtained as a solution of the Einstein 
equation with a given matter field action.

For this spacetime, Eq. (\ref{conk}) becomes
\begin{eqnarray}
\left.-C_u \sqrt{1-\dfrac{b}{r}} ~e^{\Phi} \right|_S =
\left.C_v \sqrt{1-\dfrac{b}{r}} ~e^{\Phi} \right|_S.
\end{eqnarray}
This implies that the throat candidate is the surface that satisfies $b(r)=r$ because of $C_u, C_v >0$. 
The flare-out condition 
(\ref{conr}) becomes
\begin{eqnarray}
r^a \nabla_a k|_S =\left. \dfrac{(C_u+C_v)^2 (1-b')}{4r^2} e^{-2\Phi}\right|_S >0,
\end{eqnarray}
where $b'=db(r)/dr$. This flare-out condition is satisfied if
\begin{eqnarray}
b'<1
\end{eqnarray}
on $S$. The traversability (\ref{conz}) becomes
\begin{eqnarray}
z^a \nabla_a k|_S &=&\left. \dfrac{(\alpha C_v -\beta C_u)(C_v+C_u)(1-b')}{4r^2} e^{-2\Phi}\right|_S \nonumber \\
&=&0.
\end{eqnarray}
From this equation, we see that the tangent vector of the throat orbit of 
$\alpha C_v =\beta C_u$ is timelike.

To sum up, the conditions for the wormhole are $b(r)=r$ and $b'<1$.
These conditions are the same as the well-known result of Ref. \cite{M-T}.

\subsubsection{Dynamical Ellis wormhole}\label{DEW}

The dynamical Ellis wormhole is 
sometimes investigated as a typical example \cite{M-H-C, kim}. 
The metric is given by 
\begin{eqnarray}
ds^2 & = & -dt^2 +a^2(t)[dr^2 +(r^2 +b^2) d\Omega^2_2] \nonumber \\
& = & a^2(\eta)[-d\eta^2 +dr^2 +(r^2 +b^2) d\Omega^2_2] \nonumber \\
& = & a^2(\eta)[-dudv +(r^2 +b^2) d\Omega^2_2] ,\label{metE}
\end{eqnarray}
where $a(\eta)$ is a function of the conformal time $\eta$, 
$b$ is a constant and $u=\eta -r, v=\eta +r$. 
Note that 
the metric (\ref{metE}) is not obtained as a solution of the Einstein equation 
with a given matter field action as that of the Morris-Thorne wormhole. 

For this spacetime, Eq. (\ref{conk}) becomes 
\begin{eqnarray}
C_u ( \dot{a} (r^2 +b^2) -r ) |_S =C_v ( \dot{a} (r^2 +b^2) +r ) |_S, \label{Eth}
\end{eqnarray}
where $\dot{a} =da/dt$. Since $C_u, C_v>0$, this gives us rather trivial condition 
\begin{eqnarray}
r^2 < \dot{a}^2 (r^2+b^2)^2 \label{Ellis>0}
\end{eqnarray}
at the throat candidate. The flare-out condition (\ref{conr}) 
becomes
\begin{eqnarray}
r^a \nabla_a k |_S &=&\dfrac{2C_u C_v \left[ a\ddot{a} r^2 -\dot{a}^4 (r^2 +b^2)^2 + \dot{a}^2 b^2 \right]}{a^4 \left[ \dot{a}^2 (r^2 +b^2)^2 -r^2 \right]} \bigg| _S \nonumber \\
&>&0.\label{ellisfoc}
\end{eqnarray}
Inequality (\ref{ellisfoc}) gives  
\begin{eqnarray}
f(r):=-\dot{a}^4 (r^2 +b^2)^2 + \dot{a}^2 b^2 +a\ddot{a} r^2 >0.  \label{Ellisr}
\end{eqnarray}
 
Using the 
Einstein equation $R_{\mu\nu}-Rg_{\mu\nu}/2=T_{\mu\nu}$ with the given metric (\ref{metE}), 
we compute the energy-momentum tensor $T_{\mu\nu}$. 
Then the dominant energy condition requires 
\begin{eqnarray}
&&-T^t_t -T^r_r =\dfrac{2}{a^2} (a\ddot{a} +2\dot{a}^2) \geq 0, \label{mu-p_r} \\
&&-T^t_t +T^{\theta}_{\theta} =\dfrac{2}{a^2}(-a\ddot{a} +\dot{a}^2) \geq 0, \label{mu+p_t} \\
&&-T^t_t +T^r_r =\dfrac{2}{a^2} \left( -a\ddot{a} +\dot{a}^2 -\dfrac{b^2}{(r^2 +b^2)^2} \right) \geq 0,~~ \label{mu+p_r} \\
&&-T^t_t -T^{\theta}_{\theta} =\dfrac{2}{a^2} \left( a\ddot{a} +2\dot{a}^2 -\dfrac{b^2}{(r^2 +b^2)^2} \right) \geq 0, \label{mu-p_t}
\end{eqnarray}
where $\theta$ is the angular coordinate appeared as $d\Omega_2^2=d\theta^2+\sin^2 \theta d\phi^2$. 
Inequality (\ref{mu-p_t}) gives us stronger than that obtained from inequality (\ref{mu-p_r}), which gives
\begin{equation}
a \ddot a +2 \dot a^2 \geq \frac{b^2}{(r^2+b^2)^2}. 
\end{equation}
The tightest constraint is given at $r=0$ as 
\begin{equation}
a \ddot a +2  \dot a^2 \geq b^{-2}. 
\end{equation}
In a similar way, from inequality (\ref{mu+p_r}), we have 
\begin{equation}
\dot a^2-b^{-2}  \geq a \ddot a.\label{geqddota}
\end{equation}
The above two inequalities imply  
\begin{eqnarray}
-2\dot{a}^2 +b^{-2} \leq a \ddot a \leq \dot{a}^2 -b^{-2}, \label{Edec}
\end{eqnarray}
and then we see 
\begin{eqnarray}
\dot{a}^2 b^2 \geq \dfrac{2}{3}. \label{Edec2}
\end{eqnarray}

Under the energy condition (\ref{Edec}), we can see that 
$f(r)$ has the maximum value at $r=0$.
Therefore, the condition for the existence of the region satisfying inequality (\ref{Ellisr}) is $f(0)>0$, which becomes
\begin{eqnarray}
\dot{a}^2 <b^{-2} \label{r=0}.
\end{eqnarray}
Using this, inequality (\ref{Edec}) tells us 
\begin{eqnarray}
-\dot{a}^2 <-2\dot{a}^2 +b^{-2} \leq a \ddot{a} \leq \dot{a}^2 -b^{-2} <0. \label{www}
\end{eqnarray}
Here let us suppose $a(t)$ to be proportional to $t^{\frac{2}{3(1+w)}}$, where $w$ is a 
constant. In this case, inequality (\ref{www}) implies the constraint for $w$ as 
\begin{eqnarray}
-\dfrac{1}{3} \leq w <\dfrac{1}{3}.\label{13w13}
\end{eqnarray}
We can see from inequality (\ref{Edec2}) and (\ref{r=0}) that the wormhole satisfying the 
dominant energy condition is realized in a certain time interval of the universe and 
the size is about the Hubble radius. 

Note that we did not give the equation of state like $p=w \rho$ here. In the 
current case, $w$ will be determined through the Einstein equation.
Moreover, the energy-momentum tensor derived through the Einstein equation does not have isotropic pressure. 
Therefore $w$ in inequality (\ref{13w13}) 
is not directly related to the equation of state.  

Setting $C_u=C_v=1$, we see that the throat is located at $r=0$ and the tangent of 
the throat orbit is obviously timelike.

\subsubsection{DGP wormhole}

Finally we shall consider the DGP wormhole discussed in Refs. \cite{I-S, T-S-I}. 
The DGP is one of the braneworld models and our four dimensional spacetime is realized as 
a membrane in five dimensional spacetime. In Ref. \cite{T-S-I}, Maeda et al.'s 
definition \cite{M-H-C} was employed and it turned out that the spacetime on the brane 
has the wormhole throat. Here we reconsider the brane geometry using our current definition. 

The induced metric is 
\begin{eqnarray}
ds^2 =\gamma^{-2} (r) dr^2 +r^2 \left( -d\tau ^2 +\cosh ^2 \tau d\Omega^2_2 \right),
\end{eqnarray}
where
\begin{eqnarray}
\gamma^2(r) =\dfrac{-(r^2 -2r_c^2) +\sqrt{r^4-4r_0^2 r_c^2}}{2r_c^2} \label{gamma}
\end{eqnarray}
and $r_0, r_c$ are  positive constants satisfying $r_0>r_c$. 
The range of $r$ is limited as $r \geq r_{\ast}:=\sqrt{r_0^2+r_c^2}$ 
so that $\gamma^2(r)$ is positive, and we see $0 \leq \gamma^2 (r)<1$.  

To investigate the spacetime structure in the current scheme, it is better to 
introduce  new coordinates $(T,\bar R)$ defined by $T=rh(r)\sinh \tau$ 
and $\bar R=rh(r)\cosh \tau$, where  
\begin{eqnarray}
\ln h(r)=\displaystyle \int \dfrac{1-\gamma}{\gamma r}dr. 
\end{eqnarray}
Then the metric is written as 
\begin{eqnarray}
ds^2 =h^{-2}(r)(-dT^2 +d  \bar R^2 +\bar R^2 d\Omega_2^2). 
\end{eqnarray}
Here we choose $u, v$ as $u=T-\bar R, v=T+\bar R$ and $a(u,v)=h^{-1}(r)$. 

Now we can look at Eq. (\ref{conk}) 
\begin{eqnarray}
& & \left.C_u \left[ (1-\gamma) e^\tau \cosh \tau -1 \right] \right|_S \nonumber \\
& &~~=\left.-C_v \left[ (1-\gamma) e^{-\tau} \cosh \tau -1 \right] \right|_S. \label{dgpk}
\end{eqnarray}
Because of $C_u, C_v>0$, this implies 
\begin{eqnarray}
\gamma^2 <\tanh^2 \tau. \label{g<t}
\end{eqnarray}
Note that the apparent horizon of the DGP wormhole is located at the surface satisfying 
$\gamma^2 =\tanh^2 \tau$. 

Inequality (\ref{conr}) becomes
\begin{eqnarray}
& & r^a \nabla_a k|_S \nonumber \\
& &~~=\dfrac{2C_v^2 h^2\left[ \gamma^2 (1-\gamma^2) \cosh^2 \tau +r\gamma \gamma' \sinh^2 \tau \right]}{r^2 \left\{ (1-\gamma)e^{\tau} \cosh \tau -1 \right\} ^2} \bigg| _S~~~~ \nonumber\\
&&~~>0,
\label{dgpfoc}
\end{eqnarray}
where $\gamma'=d\gamma (r) /dr$. 
Using the fact of $0\leq \gamma ^2 (r) <1$ and 
\begin{eqnarray}
r\gamma \gamma'=\dfrac{r^2}{\sqrt{r^4 -4r_0^2 r_c^2}} (1-\gamma^2)>0 \label{agg}
\end{eqnarray}
derived from Eq. (\ref{gamma}), it is easy to see that 
the flare-out condition (\ref{dgpfoc}) is always satisfied. 

The traversability (\ref{conz}) is calculated to be 
\begin{eqnarray}
&&\alpha \biggl[ C_v^2 e^{-2\tau}\left\{ r \gamma\gamma'+\left(1-\gamma^2\right)\right\} \nonumber\\
&&\qquad\qquad\qquad
\left. +C_uC_v\left\{ r \gamma\gamma'-\left(1-\gamma^2\right)\right\}\biggr] \right|_S \nonumber\\
&&=\beta \biggl[ C_u^2 e^{2\tau}\left\{ r \gamma\gamma'+\left(1-\gamma^2\right)\right\} \nonumber\\
&&\qquad\qquad\qquad
\left. +C_uC_v\left\{ r \gamma\gamma'-\left(1-\gamma^2\right)\right\}\biggr] \right|_S . \label{DGPab}
\end{eqnarray}
From Eq.~(\ref{agg}), we see $r \gamma\gamma'-\left(1-\gamma^2\right)>0$. 
This implies that $\alpha$ and $\beta$ exist and they must be positive. 
Therefore, the region satisfying inequality (\ref{g<t}) is the wormhole throat. 
This result is consistent with that in Ref.~\cite{T-S-I}.

Setting $C_u=C_v=1$, Eq. (\ref{dgpk}) is solved $r_S(\tau)$ as 
\begin{eqnarray}
r_S^2(\tau)=r_c^2(1-\tanh^4 \tau)+r_0^2(1-\tanh^4 \tau )^{-1}. 
\end{eqnarray}
This is the same with result in Ref. \cite{T-S-I}. 

Although one considers the vacuum brane, as stressed in Ref. \cite{T-S-I}, the energy conditions 
are not satisfied for the {\it effective} energy-momentum tensor computed from the four 
dimensional Einstein tensor on the brane. 

\section{summary} \label{summary}

In this paper, we proposed a new definition of the wormhole throat with the flare-out 
condition and the traversability for general cases in terms of the null expansion rate. 
This formulation is refined one of the former studies \cite{H-V1998prl, hayward1999, M-H-C}.
It can appropriately represent not only wormholes without singularities, which are mainly 
investigated in this field, but also the cosmological wormholes proposed in the recent work \cite{M-H-C}.

As a demonstration, we applied our formulation to  
several examples which include non-wormhole spacetimes too. 
As a result, we could confirm that our definition can work at least for the concrete examples 
considered here. All of our examples are spherically symmetric cases, while 
it is interesting to investigate whether in generic spacetimes our definition coincides with 
the intuitive image of wormhole. This is left for future study. 

Practically interesting objects are wormholes which we can actually pass through.  
The dynamical Ellis wormhole is that in FLRW universe without violating any exotic matters, 
and thus it could exist in our Universe. However, it is too large. Because of the similar size 
to the Hubble radius, even if it exists, it is not observed as a compact object but rather 
affects to the cosmological scale physics. For actual use, small wormholes are fascinating, but 
it seems hard or impossible to construct such wormholes without violating the energy condition.

\begin{acknowledgments}
Y.~T. thanks Professor Hiroaki Kanno for his continuous encouragement.
K.~I. is supported by Taiwan National Science Council under Project No. NSC101-2811-M-002-103. 
T.~S. is supported by Grant-Aid for Scientific Research from Ministry of 
Education, Science, Sports and Culture of Japan (No.25610055). 

\end{acknowledgments}



\end{document}